\documentclass[10pt,superscriptaddress,twocolumn,amsmath,amssymb,aps,prl,showpacs]{revtex4-1}
\usepackage{mathrsfs}
\usepackage{graphicx}
\usepackage{dcolumn}
\usepackage{bm}
\usepackage[title,titletoc,header]{appendix}
\usepackage{amssymb}
\usepackage{amsmath}
\usepackage{mathrsfs}
\usepackage{amsmath}
\usepackage{subfigure}
\usepackage{float}
\usepackage[title,titletoc,header]{appendix}
\usepackage{graphicx}

\newcommand{\be}{\begin{equation}}
\newcommand{\ee}{\end{equation}}
\newcommand{\bea}{\begin{eqnarray}}
\newcommand{\eea}{\end{eqnarray}}

\begin{document}
\title{Collective Modes in a Two-band Superfluid of Ultracold Alkaline-earth Atoms Close to an Orbital Feshbach Resonance}
\author{Yi-Cai Zhang}
\affiliation{Department of Physics and Center of Theoretical and
Computational Physics, The University of Hong Kong, Hong Kong, China}

\author{Shanshan Ding}
\affiliation{Department of Physics and Center of Theoretical and
Computational Physics, The University of Hong Kong, Hong Kong, China}

\author{Shizhong Zhang}
\affiliation{Department of Physics and Center of Theoretical and
Computational Physics, The University of Hong Kong, Hong Kong, China}

\date{\today}
\begin{abstract}
We discuss the collective modes in an alkaline-earth Fermi gas close to an orbital Feshbach resonance. Unlike the usual Feshbach resonance, the orbital Feshbach resonance in alkaline-earth atoms realizes a two-band superfluid system where the fermionic nature of both the open and the closed channel has to be taken into account. We show that apart from the usual Anderson-Bogoliubov mode which corresponds to the oscillation of total density, there also appears the long-sought Leggett mode corresponding to the oscillation of relative density between the two channels. The existence of the phonon and the Leggett modes and their evolution are discussed in detail. We show how these collective modes are reflected in the density response of the system.
\end{abstract}
\maketitle

{\em Introduction}. Recent theoretical proposal~\cite{Zhaihui} and the subsequent experimental confirmations~\cite{Fallani,Folling} of the so-called orbital Feshbach resonance (OFR) in alkaline-earth atoms have opened a new avenue to investigate strongly interacting Fermi gases. A prominent feature of the alkaline-earth atoms, such as Sr~\cite{Ye}, Yb~\cite{Cappellini,Scazza,Taie}, is that there are two clock states, the electronic $s$- and $p$-states, both of which have zero electronic angular momentum $J=0$. As a result, there is no hyperfine coupling between the electronic and nuclear spins. The inter-atomic interactions which depend (primarily) on the electronic configurations thus become independent of nuclear spins. This realizes the so-called $SU(N)$ symmetries, where $N$ is the number of nuclear spin components~\cite{Wu2003,Wu2005,Hung2011,Wang2014,Gorshkov,Cazalilla,Pagano,Zhou2016}. The OFR relies on the atoms residing on both the $s$ and $p$-states which possess slight different Land\'{e} g-factors~\cite{Lurio,Boyd}, thus allowing tuning by external magnetic field.

Unlike the usual Feshbach resonance, where the closed channel bound state responsible for the resonance has a binding energy relative to its scattering threshold much larger than the Fermi energy of the system, OFR in turn relies on the existence of a shallow two-body bound state whose binding energy is of order of the Fermi energy, and thus brings in new features that are not encountered before. In particular, it is no longer admissible to treat the closed channel state as a single boson without taking into account of its proper internal dynamics. One is thus lead to a situation much the same as a two-band superconductor, where the appropriate picture is that two Fermi surfaces (including both spin) intersect the chemical potential \cite{Iskin2005,Klimin,Xu2016}. We show that this new situation leads to the appearance of new collective modes and in particular, the long-sought Leggett mode in the two-band superconductor. The dependences of the collective modes on tuning parameters are analyzed in detail. We also calculate the dynamic structure factor of the system and identify the appearance of the Leggett mode in the oscillations of the relative density.

{\em The Model Hamiltonian}. Let us denote the electronic ground $s$-state as $g$ and the excited $p$-state as $e$. Then the Hamiltonian of alkaline-earth fermions close to an orbital Feshbach resonance is given by $H=H_{\rm o}+H_{\rm c}+H_{\rm int}$ \cite{Zhaihui}, where (setting $m=\hbar=1$)
\begin{align}
 &H_{\rm o}=\sum_{\bf k} \epsilon_{\bf k}(c^\dagger_{g\downarrow, {\bf k}}c_{g\downarrow, {\bf k}}+c^\dagger_{e\uparrow, {\bf k}}c_{e\uparrow, {\bf k}}),\\
 &H_{\rm c}=\sum_{\bf k} (\epsilon_{\bf k}+\delta/2)(c^{\dagger}_{g\uparrow, {\bf k}}c_{g\uparrow, {\bf k}}+c^{\dagger}_{e\downarrow,{\bf k}}c_{e\downarrow, {\bf k}}),\\
 &H_{\rm int}=-\frac{g_+}{2}A^{\dagger}_{+}A_+-\frac{g_-}{2}A^{\dagger}_{-}A_-.
\end{align}
Here $H_{\rm o} (H_{\rm c})$ are the single-particle Hamiltonian for the open (closed) channels, $\epsilon_{\bf k}=k^2/2-\mu$ with $\mu$ being the chemical potential. $\delta$ is the detuning between the open and close channels and can be tuned by external magnetic field. $A_+=\sum_{\bf k} (c_{e\downarrow,{\bf k}}c_{g\uparrow, -{\bf k}}-c_{e\uparrow, {\bf k}}c_{g\downarrow, -{\bf k}})$ and $A_-=\sum_{\bf k}(c_{e\downarrow, {\bf k}}c_{g\uparrow, -{\bf k}}+c_{e\uparrow, {\bf k}}c_{g\downarrow, -{\bf k}})$. Here $g_{+(-)}>0$ are the coupling constants indicating the strength of attractive interaction for the singlet (triplet) orbital channel and related to the respective scattering lengths $a_{+/-}$ by the usual renormalization conditions: $1/g_{+/-}=-1/4\pi a_{+/-}+\sum_{\bf k}1/k^2$.

To investigate the paring structure of Fermi gases close to OFR, we define the order parameters $\Delta_{+/-}\equiv -g_{+/-}\langle A_{+/-}\rangle/2$, or more conveniently, $\Delta_{\rm o}\equiv\Delta_--\Delta_+$ and $\Delta_{\rm c}\equiv\Delta_-+\Delta_+$ for the pairing strength in the open and closed channel, respectively. Introducing the Grassman variable in the Nambu notation, $\bar{\Psi}=(\bar{\psi}_{g\downarrow},\psi_{e\uparrow},\bar{\psi}_{g \uparrow},\psi_{e\downarrow})$, the partition function of the system can be written as, after a Hubbard-Stratonovich transformation, $\mathcal{Z}\equiv e^{-\beta \Omega}=\int D\bar{\Psi}D{\Psi}D\bar{\Delta}_{\rm o}D\Delta_{\rm o}D\bar{\Delta}_{\rm c}D\Delta_{\rm c}\exp(-S)$, where the effective action
\begin{align}
S=-\int dx \left[\bar{\Psi}G^{-1}\Psi-\frac{2|\Delta_+(x)|^2}{g_+}-\frac{2|\Delta_-(x)|^2}{g_-}\right],
\end{align}
here $x=\{\tau,\vec{r}\}$, $\int dx\equiv \int_{0}^{\beta}d\tau \int d^3 \vec{r}$ with $\beta=1/k_{\rm B}T$ being the inverse temperature and $\Omega$ is thermodynamical potential. In this work, we focus on the zero-temperature physics and take the limit of $\beta\rightarrow\infty$ at the end of calculation.
The inverse Green function $G^{-1}$ is given by
\begin{align}
G^{-1}(x,x')=\delta(x-x')\left(\begin{array}{cc}
   G^{-1}_{\rm o}    &  0\\
   0 &   G^{-1}_{\rm c}\\
\end{array}\right)
\end{align}
with the matrix Green functions $ G^{-1}_{\rm o, c}$ given by
\begin{align}
G^{-1}_{\rm o}(x)=\left(
 \begin{array}{cccc}
   -\partial_\tau+{\nabla^2}/{2}+\mu  &    \Delta_{\rm o}(x)    \\
   \Delta^*_{\rm o}(x)     &   -\partial_\tau-{\nabla^2}/{2}-\mu       \\
\end{array}\right)
\end{align}
and
\begin{align}
G^{-1}_{\rm c}(x)=\left(
 \begin{array}{cccc}
   -\partial_\tau+\frac{\nabla^2}{2}-\frac{\delta}{2}+\mu  &    \Delta_{\rm c}(x)    \\
   \Delta^*_{\rm c}(x)     &   -\partial_\tau-\frac{\nabla^2}{2}+\frac{\delta}{2}-\mu       \\
\end{array}\right).
\end{align}
Integrating over the fermionic field, one finds the effective action $S_{\Delta}$ in terms of $\Delta_{\rm o,c}$, given by
\begin{align}
S_{\Delta}=\int dx \left[\frac{2|\Delta_+(x)|^2}{g_+}+\frac{2|\Delta_-(x)|^2}{g_-}\right]-{\rm Tr}\ln G^{-1}(x,x')
\end{align}
In the following,  we choose the parameter such that $k_Fa_{+}=1$, and investigate the evolution of collective mode as one changes $a_-$. Here $k_F$ is determined by the density through $n=k_{F}^3/(3\pi^2)$.

\begin{figure}
\begin{center}
\includegraphics[width=\columnwidth]{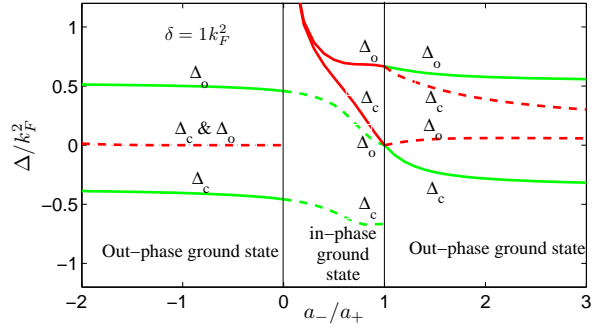}
\end{center}
\caption{(Color online) The order parameters for the two band superfluid close to a orbital Feshbach resonance. Depending on the values of $a_-/a_+$, there are in general two types of solution to the mean field equation. The red lines indicate in-phase solution while the green lines are for the out-of-phase solution. The solid lines indicate stable solution and the dashed lines are for the saddle point solution for specific parameter range of $a_-/a_+$. We choose the parameters $\delta=k_{F}^2$.}
\label{fig1}
\end{figure}

{\em Mean field solution and its stability.} Within mean field approximation, the optimal values of the order parameters $\Delta_{\rm o,c}$ are given by the saddle point equations: $\partial \Omega/\partial\Delta_{\rm o,c}=0$. On the other hand, the chemical potential is determined by $\partial \Omega/\partial \mu=-n$ with $n$ the average density. It turns out that, depending on the value of $a_-/a_+$, there are two types of solutions: the true ground state and the metastable state. For $0<a_-/a_+<1$, the ground state corresponds to the case when two order parameters are in phase: $\Delta_{\rm c}\Delta_{\rm o}>0$, while the out-of-phase solution $\Delta_{\rm c}\Delta_{\rm o}<0$ is the metastable state. On the other hand, for $a_-/a_+<0$ or $a_-/a_+>1$, the out-of-phase solution is the true ground state, while the in-phase solution is the metastable solution; see Figure \ref{fig1}.

To understand the various mean field solutions, we assume that $a_{+}>0$. The coupling energy between the two channels within mean-field approximation is given by $\langle V_{\rm oc}\rangle=2(1/a_{+}-1/a_{-})|\Delta_{\rm o}\Delta_{\rm c}|\cos(\theta_{\rm o}-\theta_{\rm c})$, where $\theta_{\rm o,c}$ are the phases of the order parameters $\Delta_{\rm o,c}$. When $0<a_{-}<a_{+}$, the bound state in the triplet potential (determined by $a_{-}$) has lower energy and as a result, the triplet pairing order parameter $\Delta_{-}=\Delta_{\rm o}+\Delta_{\rm c}$ is dominant over the singlet component $\Delta_{+}$. As a result, in order to minimize the coupling energy $\langle V_{\rm oc}\rangle$, $\Delta_{\rm o}\Delta_{\rm c}>0$ and in-phase solution is the ground state. On the other hand, when $a_->a_+$ or $a_-<0$, the singlet channel is the dominant pairing component and the out-of-phase solution $\Delta_{\rm o}\Delta_{\rm c}<0$ is the true ground state. Hence, it is expected that as one changes $a_-$, the ground state properties will change from a singlet dominant pairing to a triplet one. At the transition when $a_-=a_+$, the the open and closed channels decouple (the coupling constant $\propto a_- -a_+$)~\cite{Zhaihui}. The two channels become independent from each other, so the the out-phase and in-phase solutions are degenerate; See Figure 1.  The experimental interaction parameters near the orbital-Feshbach resonance of $^{173}$ Yb atomic gas are  $a_+\sim 1900a_0$ and $a_-\sim200a_0$, giving $a_-/a_+\sim0.1$~\cite{Fallani,Folling}. So the ground state of  system is in-phase  solution of $\Delta_{\rm c}\Delta_{\rm o}>0$ close to OFR.  In this work, we shall mainly focus on the true ground state and its associated collective excitations for different parameters  of  $a_-/a_+$ and the detuning $\delta$.

{\em Collective excitations}. To investigate the fluctuation around the mean field solutions, we write $\Delta_{\rm o,c}(x)=\Delta_{\rm o,c}+\eta_{\rm o,c}(x)$. Transforming to the momentum and Matsubara frequency space, we can write the inverse Green function as $G^{-1}(k,k')=G_0^{-1}(k)\delta(k-k')+K(k,k')$, where
\begin{align}
G_0^{-1}(k)\!\!=\!\!
\left(\!\!
 \begin{array}{cccc}
   i\omega_n-\epsilon^{o}_{k}  &   \Delta_o    &   0    &  0\\
  \Delta^*_o   &      i\omega_n+\epsilon^{o}_{k}    &     0   &  0\\
   0    &    0    &    i\omega_n-\epsilon^{c}_{k} &  \Delta_c\\
   0    &     0    &   \Delta^*_c   &   i\omega_n+\epsilon^{c}_{k}
\end{array}\!\!\right)
\label{G0}
\end{align}
and the matrix $K(k,k')$ is given by
\begin{align}
K (k,k')=\left(
 \begin{array}{cccc}
   0  &  \eta_{\rm o}(-q)    &    0    &  0\\
 \eta_{\rm o}^{*}(q)    &   0     &     0   &   0 \\
    0     &    0    &    0   &  \eta_{\rm c}(-q)\\
   0    &     0     &   \eta_{\rm c}^{*}(q)  &   0
\end{array}\right).
\end{align}
Here $k=\{i\omega_n, {\bf k}\}$ and $\delta(k-k')\equiv\delta^3({\bf k}-{\bf k}')\delta_{nn'}$. $\epsilon^{\rm o}_{\bf k}=k^2/2-\mu$ and $\epsilon^{\rm c}_{\bf k}=k^2/2+\delta/2-\mu$ are the kinetic energy of the open and closed channel measured from the chemical potential $\mu$. The four-momentum transfer $q=k-k'\equiv \{{\bf k}-{\bf k}', i\omega_n-i\omega_{n'}\}$. The fluctuation contribution to the effective action then can be written in the usual quadratic form $S_\eta=\frac{1}{2}\sum_{q}\bar{\eta}_qM(q)\eta_q$ where the fluctuation matrix is given by \cite{Diener}
\begin{align}
M=\left(
 \begin{array}{cccc}
   M^{\rm o}_{11}(q)  &  M^{\rm o}_{12}(q)    &   dg    &  0\\
  M^{\rm o}_{21}(q)    &      M^{\rm o}_{22}(q)    &     0   &  dg\\
   dg    &    0    &    M^{\rm c}_{11}(q)  &  M^{\rm c}_{12}(q)\\
  0    &     dg    &   M^{\rm c}_{21}(q)   &  M^{\rm c}_{22}(q)
\end{array}\right).
\end{align}
Here $\bar{\eta}_q =[\eta^{*}_{\rm o}(q),\eta_{\rm o}(-q),\eta^{*}_{\rm c}(q),\eta_{\rm c}(-q)]$ and  $dg=1/2(1/g_{-}-1/g_{+})=-1/8\pi({1}/{a_{-}}-{1}/{a_{+}})$. The various $M$'s are given by
\begin{align}
M^{\rm o}_{11}(q)&\!\!= \!\!\frac{1}{\beta}\sum_{k}G_{\rm o11}(k+q)G_{\rm o22}(k)\!\!+\!\!\frac{1}{2}\left[\frac{1}{g_+}+\frac{1}{g_-}\right],\\
M^{\rm o}_{12}(q)&=\frac{1}{\beta}\sum_{k,n'}G_{\rm o12}(k+q)G_{\rm o12}(k),\\
M^{\rm o}_{21}(q)&=M^{\rm o}_{12}(q),\\
M^{\rm o}_{22}(q)&=M^{\rm o}_{11}(-q),
\end{align}
and similarly for the closed channel.
The collective modes are given by the zeros of determinant ${\rm Det}|M({\bf q},\omega+i0^+)|=0$~\cite{Melo}.

\begin{figure}
\begin{center}
\includegraphics[width=\columnwidth]{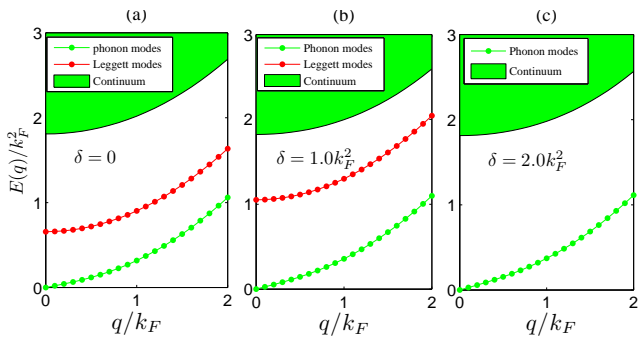}
\end{center}
\caption{(Color online) The dispersions of the collective excitations for different detuning for the ground state with $a_{-}/a_{+}=0.8$. (a) For $\delta=0$, both Leggett and phonon modes are well defined and below the quasi-particle continuum. (b) As detuning increased to $\delta=1k_{F}^{2}$, the gap for the Leggett mode at ${\bf q}=0$ is pushed upwards. (c) For $\delta=2k_{F}^{2}$, only phonon mode remains and the Leggett mode merges into the continuum.}
\label{fig2}
\end{figure}

In Figure \ref{fig2} (a-c), we show the excitation spectrums for three different detunings $\delta=0,1,2$ (in unit of $k_F^2$) and for fixed values of $a_{-}/a_{+}=0.8$. Because of the existence of two bands, in addition to the usual Goldstone (or Anderson-Bogoliubov) modes (green lines in Fig.2), which corresponds to the oscillation of total density, there appears additional Leggett mode (red lines), which corresponds to the oscillation of the relative densities of the two bands \cite{Leggett}.  With increasing detuning $\delta$, the Leggett modes gradually merge into the two quasi-particle continuum and are heavily damped.

One can understand the appearance of the Leggett mode as follows. In the absence of inter-channel coupling ($a_-=a_+$), the pairing occurs in the closed and the open channel independently. As a result, two phonon modes appear corresponding to the independent density fluctuations in the two channels. As the inter-channel coupling ($a_-\neq a_+$) is turned on, there still remains a phonon mode corresponding to the total density fluctuations, while the fluctuation of relative densities in the two channels acquires a non-zero gap due to inter-channel coupling. In Figure 3(b), we calculated the variations of gap of the Leggett mode $\omega_L({\bf q}=0)$ with the scattering length $a_{-}/a_{+}$ in ground states. As expected, near the phase transition point ($a_{-}/a_{+}=1$) where inter-channel coupling vanishes, $\omega_L({\bf q}=0)\to 0$.

\begin{figure}[h]
\begin{center}
\includegraphics[width=\columnwidth]{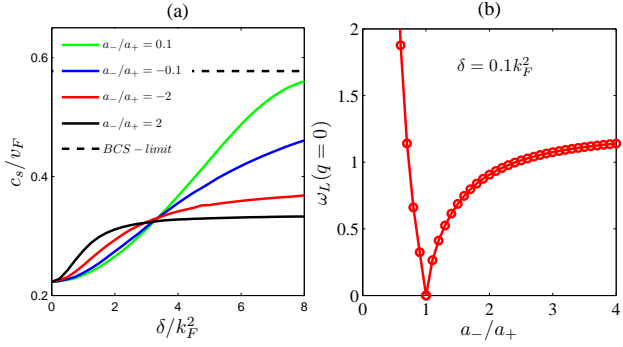}
\end{center}
\caption{(Color online) (a) The evolution of sound velocity $c_s$ with detuning $\delta$ for various parameter $a_{-}/a_+$ for out-phase solution $\Delta_{\rm o} \Delta_{\rm c}<0$. The dashed line indicates the limiting value with $c_s={1}/{\sqrt{3}}v_F$ in the BCS limit. (b) The evolution of gap for the Leggett mode with $a_{-}/a_{+}$ for $\delta=0.1k_{F}^{2}$ in the ground states. Note that the gap closes at $a_{-}/a_{+}=1$ where the coupling between the two channels vanishes.}
\label{fig3}
\end{figure}

Figure \ref{fig3} (a)  shows the variations of sound velocity as a function of  $\delta$ with   different $a_-/a_+$  for out-phase solution ($\Delta_{\rm o} \Delta_{\rm c}<0$).
When the detuning $0<a_-/a_+<1$, the system is  metastable (a saddle point of energy), however there still exist well-defined phonon modes (green line) in Figure 3(a). This is because for $q\rightarrow0$, the phonon modes correspond to in-phase density oscillation of two channels that is along the direction along which the energy increases. In addition, with increasing of $\delta$, the system enters to the BCS limit \cite{Zhaihui}. Consequently,  sound velocity also saturates to its BCS limit value $c_s=v_F/\sqrt{3}$ (green line)~\cite{Diener}.
For other several values of $a_-/a_+=-0.1,-2, 2$, the system is in its true ground state and the sound velocity increases with increasing of detuning $\delta$.
 When $\delta=0$, the Hamiltonian can be written as two independent (singlet and triplet orbital) channel  Hamiltonian by reorganizing the single-particle Hamiltonian  of Eq. (1).
 The out-phase solution would satisfy $\Delta_{\rm o}=-\Delta_{\rm c}$. Hence only the singlet orbital paring $\Delta_+=(\Delta_{\rm c}-\Delta_{\rm o})/2$ occurs, while the triplet orbital paring $\Delta_-=(\Delta_{\rm c}+\Delta_{\rm o})/2$ vanishes. Consequently the thermodynamical potential does not depend on $a_-$ and sound velocities for different $a_-/a_+$ are same. In addition, for fixed $\delta$, sound velocity increases when $a_-/a_+$ varies from $-0.1$ to $2$ when detuning is small while deceases for larger $\delta$.

{\em Dynamical structure factor}. As a direct probe of the Leggett mode, one can use the Bragg spectroscopy which is related to the density fluctuations of the system \cite{Vale}. We define the density correlation function matrix as
\begin{align}
\chi(q)\equiv\left(
 \begin{array}{cc}
   \langle\delta\rho_{\rm o}(-q)\delta\rho_{\rm o}(q)\rangle &   \langle\delta\rho_{\rm o}(-q)\delta\rho_{\rm c}(q)\rangle\\
  \langle\delta\rho_{\rm c}(-q)\delta\rho_{\rm o}(q)\rangle &   \langle\delta\rho_{\rm c}(-q)\delta\rho_{\rm c}(q)\rangle \\
\end{array}\right),
\end{align}
where $\delta\rho_{\rm o,c}(q)$ are the density fluctuation operators in the open and closed channel. The explicit form of $\chi(q)$ in terms of the various components of $G_0$ [Eq. \ref{G0}] can be arranged as $\chi(q)=a-bM^{-1}c$ with $a,b,c$ given by
\begin{align}
a &=\left(\begin{array}{cc}
   \pi^{\rm o}_{11}(q) &   0   \\
  0 &   \pi^{\rm c}_{11}(q) \\
\end{array}\right),\\
b &=\left(\begin{array}{cccc}
  \pi^{\rm o}_{12}(q) &   \pi^{\rm o}_{13}(q) & 0  &0 \\
  0 &  0&  \pi^{\rm c}_{12}(q) &   \pi^{\rm c}_{13}(q)\\
\end{array}\right),\\
c&=\left(\begin{array}{ccc}
   \pi^{\rm o}_{21}(q)  &0 \\
  \pi^{\rm o}_{31}(q)  &0 \\
  0 &\pi^{\rm c}_{21}(q)  \\
   0 &\pi^{\rm c}_{31}(q)  \\
\end{array}\right).
\end{align}
The polarization operators are given by
\begin{align}
\pi^{\rm o}_{11}(q)&= \frac{2}{\beta}\sum_{k}(G_{\rm o11}(k+q)G_{\rm o22}(k)-G_{\rm o12}(k+q)G_{\rm o21}(k)),\\
\pi^{\rm o}_{12}(q)&=\pi^{\rm o}_{21}(q)= \frac{2}{\beta}\sum_{k}G_{\rm o21}(k+q)G_{\rm o22}(k),\\
\pi^{\rm o}_{13}(q)&=\pi^{\rm o}_{31}(q)= \frac{2}{\beta}\sum_{k}G_{\rm o22}(k+q)G_{\rm o12}(k)
\end{align}
and similarly for the closed channel components. The dynamical structure factor for the total density fluctuation is given by ($i\omega_n\to\omega+0^+$)
\begin{align}
S_{\rm total}(\omega)=-\frac{1}{\pi}{\rm Im}(\chi_{11}+\chi_{12}+\chi_{21}+\chi_{22})
\end{align}
and that for the relative density oscillation is given by
\begin{align}
S_{\rm rel}(\omega)=-\frac{1}{\pi}{\rm Im}(\chi_{11}-\chi_{12}-\chi_{21}+\chi_{22}).
\end{align}
The dynamical structure  factor $S_{\rm total}$ satisfies the famous f-sum rule $\int d\omega \omega S_{\rm total}({\bf q},\omega)[[\rho_{\bf q},H],\rho_{-{\bf q}}]/2=Nq^2/2m$, where $\rho_{\bf q}=\rho^{\rm o}_{\bf q}+\rho^{\rm c}_{\bf q}$ is the total density fluctuation operator and $N=N_{\rm o}+N_{\rm c}$ is the total number of fermions. Defining the relative density fluctuation operator $m_{\bf q}\equiv\rho^{\rm o}_{\bf q}-\rho^{\rm c}_{\bf q}$, one can show that, in the limit when $q\to 0$ (or more precisely $qr_0\ll 1$, where $r_0$ is the range of actual inter-atomic potential)
\begin{align}
\int d\omega \omega S_{\rm rel}({\bf q},\omega)=\frac{1}{2}[[m_{\bf q},H],m_{-\bf q}]=\frac{Nq^2}{2m}-8\langle V_{\rm oc}\rangle.
\end{align}
$\langle V_{\rm oc}\rangle$ is the coupling energy between the open and the closed channels. This is an exact relation and does not depend on the quantum state. For true ground state, the coupling energy $\langle V_{\rm oc}\rangle<0$, so the contribution to the sum rule is positive; while for saddle point solution ($\langle V_{\rm oc}\rangle>0$), the contribution is negative. When $q=0$, the sum rule for the relative density fluctuation gives directly the coupling energy $\langle V_{\rm oc}\rangle$, which quantifies the correlation between the open and the closed channels.

\begin{figure}
\begin{center}
\includegraphics[width=\columnwidth]{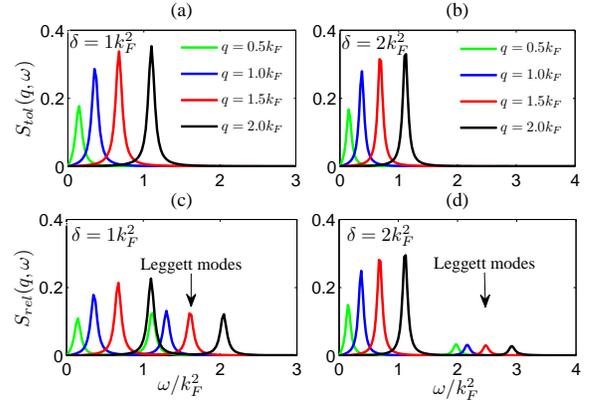}
\end{center}
\caption{(Color online) The dynamical structure factors for the total density (a,b) and the relative density (c,d) for $\delta=1k_{F}^{2}$ and $\delta=2k_{F}^{2}$ with $a_-/a_+=0.8$. (a,b) only phonon mode is present in $S_{\rm total}({\bf q},\omega)$ and its frequency increases with wave vector $q$. (c,d) $S_{\rm rel}({\bf q},\omega)$ features peaks corresponding to Leggett mode at relative higher energies. }
\label{fig4}
\end{figure}

Fig. \ref{fig4} shows dynamical structure factors $S_{\rm total}({\bf q},\omega)$ and $S_{\rm rel}({\bf q},\omega)$ for $a_{-}/a_{+}=0.8$, $\delta=1k_{F}^{2}$ [Figure \ref{fig4} (a,c)] where the Leggett mode is well defined and $\delta=2k_{F}^{2}$ [Figure \ref{fig4} (b,d)] where the Leggett mode is within the two-particle continuum and  damped [see Fig.\ref{fig2}(c)]. In Fig.\ref{fig4}(a,b), $S_{\rm total}({\bf q},\omega)$ only features a low frequency peaks corresponding to phonon excitations while the Leggett mode is absent. For $S_{\rm rel}({\bf q},\omega)$, in addition to the phonon modes, there appear high frequency peaks corresponding to well defined [Fig.\ref{fig4}(c)] and damped [Fig.\ref{fig4}(d)] Leggett mode. Comparing Fig.\ref{fig4} (c) with Fig.\ref{fig4} (d), the spectral weight of Leggett mode diminishes after merging into the continuum. Investigations of the Leggett modes in multi-band superconductor have already attracted intensive interests~ \cite{Ichioka,Sharapov,Tanaka2002,XiaoHu2012,Marciani,Tanaka2015}. It is only until very recently years that some evidences of its existence have been observed experimentally in multi-band superconductors MgB$_2$, by tunneling spectroscopy techniques~\cite{Ponomarev2004}, Raman spectroscopy~\cite{Blumberg2007}, and angle-resolved photoemission spectroscopy~\cite{Mou2015}. In the case of cold atom system, the Leggett modes have so far not been observed experimentally. One can expect that with the realization of a two-band superfluid Fermi gas, the Leggett modes would appear as resonance peaks in the Bragg spectroscopy.

{\em Conclusions}.
Motivated by experimental realization of orbital Feshbach resonance,  we have investigated the collective excitations of two-band superfluid near the orbital Feshbach resonance. We identified the existence of the (damped/undamped) Leggett mode in the relative density response of the system, which can be measured using Bragg spectroscopy. The vanishing of the gap of the Leggett mode can be used further to identify the quantum phase transition from the singlet to the triplet orbital pairing.

{\em Acknowledgement}. We thank Hui Zhai for useful discussions. This work is supported by Hong Kong Research Grants Council, General Research Fund, HKU 17306414, CRF, HKUST3/CRF/13G, and the Croucher Foundation under the Croucher Innovation Award.

{\em Note}: In the preparation of the work, we beware of  two relevant works appearing \cite{Iskin,Heliangyi}.

\end{document}